\documentclass[10pt,oneside,onecolumn,preprint,number,sort,compress]{elsarticle}
\pdfoutput=1
\usepackage{amsmath}
\usepackage{amssymb}
\usepackage{subfigure}
\usepackage{hyperref}
\usepackage{mathrsfs}
\usepackage{amsxtra}
\usepackage{mathrsfs}
\usepackage{amsfonts}
\usepackage{epsfig}
\usepackage{dsfont}
\usepackage{dcolumn}
\usepackage{bm}
\usepackage{graphicx}
\usepackage{epstopdf}
\usepackage{txfonts}

\begin{document}

\begin{frontmatter}
\title{Trapping Penguins with Entangled B Mesons}
\author[val]{Ryan Dadisman}\ead{jrdadi2@g.uky.edu}
\author[val]{Susan Gardner}\ead{gardner@pa.uky.edu}
\author[val]{Xinshuai Yan}\ead{xya226@g.uky.edu}
\address{Department of Physics and Astronomy, University of Kentucky,
Lexington, Kentucky 40506-0055 USA}

\begin{abstract}
The first direct observation of time-reversal (T)
violation in the $B\overline{B}$ system
has been reported by the BaBar collaboration, employing the method of 
Ba${\tilde{\rm n}}$uls and 
Bernab{\'e}u.
Given this, we generalize their analysis of the time-dependent T-violating asymmetry ($A_{T}$)
to consider different choices of CP tags for which the dominant 
amplitudes have
the same weak phase. 
As one application, we find 
that it is possible to measure
departures from the universality of $\sin(2\beta)$ directly. If $\sin(2\beta)$ is universal, as in 
the Standard Model, 
the method permits the direct determination of penguin effects in these channels. 
Our method, although no longer a strict test of T, can yield 
tests of the $\sin(2\beta)$ universality, or, alternatively, of penguin effects, of 
much improved precision even with existing data sets. 
\end{abstract}
\end{frontmatter}

\section{Introduction}
\label{sec:Introduction}
A goal of $B-$physics is to study the nature of CP violation and to discern,
ultimately, whether sources of CP violation exist beyond that of the Standard Model (SM).
This means 
the weak phases
associated
with various decays are measured to test whether they fit the SM pattern or not. Thus far
such searches have proven nil, noting, e.g., Ref.~\cite{Laiho:2009eu} and
its update in Ref.~\cite{Brambilla:2014jmp},
and it is of interest to carry these tests to
higher precision.
For example, in the SM the CP asymmetries associated with
the quark decays $b\to c\bar{c}s$, $b\to c\bar{c} d$, and $b\to s\overline{s}s$
measure $\sin(2\beta)$, 
up to penguin contributions and new physics
in the decay amplitudes~\cite{Grossman:1996ke}.\footnote{Recall
$\beta \equiv\hbox{arg}[- V^{}_{cd} V_{cb}^\ast/(V^{}_{td} V_{tb}^\ast)]$, where
$V_{ij}$ is an element of the Cabibbo-Kobayashi-Maskawa (CKM) matrix. In this paper
we use ``penguin contributions'' to connote all wrong 
phase contributions to the decay amplitude.} Measurements of the
time-dependent asymmetry in the penguin mode $B\to \phi K_S$ ($b\to s{\bar s} s$)
and others
are statistics limited, and follow-up studies
are planned at Belle-II~\cite{Abe:2010gxa} 
A compilation
of existing measurements can be found in Ref.~\cite{Bevan:2014iga}.
Improved tests of weak-phase universality, notably that of $\sin(2\beta)$,
using the usual measurement of time-dependent CP asymmetries will require 
experiments at new facilities. 
In this paper, we propose a more accessible way to sharpen these tests
by determining effective weak-phase differences through
a single asymmetry measurement; thus an improved test can come from existing data sets. 

The BaBar collaboration has observed direct T violation~\cite{Lees:2012uka}
by exploiting the quantum entanglement of the $B {\bar B}$ mesons
produced in $\Upsilon(4S)$ decays, as long 
familiar from other contexts~\cite{Bigi:1981qs,Okun:1975di,Bigi:1986dp,Atwood:2002ak}. 
That is, because the $\Upsilon(4S)$ state has definite flavor and CP,
the flavor- or CP-state of a
$B$ meson can be determined, or ``tagged,'' at a time $t$ by measuring the
decay of the other $B$ meson at that instant. In a seminal paper, 
Ba${\tilde{\rm n}}$uls and Bernab{\'e}u showed that
by selecting suitable combinations of flavor and CP tags 
of the $B$-mesons 
in the entangled pair, CP, T, and CPT asymmetries~\cite{Banuls:1999aj} can all be constructed. 
Consequently, BaBar uses the final
states $J/\Psi K_L$ ($\text{CP}=+$) and $J/\Psi K_S$ ($\text{CP}=-$)
as CP tags and the sign of the charged lepton in $\ell^{\pm}X$ decay as a
flavor tag. Thus by employing either 
flavor or CP tagging they are able to
form a time-dependent asymmetry $A_T$, such as
$A_T = (\Gamma(B^0 \to B_+) - \Gamma(B_+ \to B_0))/(\Gamma(B^0 \to B_+)
+ \Gamma(B_+ \to B_0))$, where $B_\pm$ denotes a state with
$\text{CP}=\pm$~\cite{Banuls:1999aj,Banuls:2000ki,Bernabeu:2012ab,Lees:2012uka}.
Thus if the rates of 
$B^0 \to B_+$ and $B_+ \to B^0$ are not the same, i.e., not in ``detailed balance,''
then time-reversal symmetry is broken. 
BaBar measures the T-violating parameters 
$\Delta S_{T}^{+} = -1.37 \pm 0.14_{\rm stat} \pm 0.06_{\rm syst}$ and
$\Delta S_{T}^{-} = 1.17 \pm 0.18_{\rm stat} \pm 0.11_{\rm syst}$, so that both
measurements exceed discovery significance, 
and reports observing 
T violation with an effective  significance of $14\sigma$~\cite{Lees:2012uka}. 
Previously a failure of detailed balance was reported in $K^0 \leftrightarrow {\bar K}^0$
transitions by CPLEAR~\cite{Angelopoulos:1998dv}, but the concomitant claim
of direct T violation of 
$\langle A_{T}^{exp} \rangle = (6.6 \pm 1.3_{\rm stat} \pm 1.0_{\rm syst}) \times 10^{-3}$ 
is only of $4\sigma$ significance if statistical and systematic errors are combined
in quadrature. Moreover, the interpretation of the experiment as a test of T 
has been criticized~\cite{Wolfenstein:1999re,Wolfenstein:2000ju}.
In the case of the concept~\cite{Banuls:1999aj,Banuls:2000ki} employed by the 
BaBar experiment~\cite{Lees:2012uka},
the use of entanglement with distinct kinds of tags
allows the reservations~\cite{Wolfenstein:1999re,Wolfenstein:2000ju}
levied against the CPLEAR experiment 
to be set to rest~\cite{Alvarez:2006nk,Quinn:2009zza,Brambilla:2014jmp}.  

Nevertheless, there has been discussion of the conditions under which a 
measured non-zero value of $A_T$ proves that time-reversal symmetry is broken. 
Generally, the existence of penguins complicate the interpretation of these measurements
as tests of T (or of CPT), 
though in the specific final states studied by BaBar~\cite{Lees:2012uka} 
$A_T$ is a true test of T irrespective of penguin effects in the 
$B$-meson decay~\cite{Applebaum:2013wxa}.
Direct CP violation in the CP tag, however, which is possible if 
$K_{S,L}$ are reconstructed through their hadronic decays, 
also causes the interpretation of $A_T$ as a
test of T to fail --- this has also been
noted by Ref.~\cite{Bernabeu:2012nu} in an analogous
study of $K{\bar K}$ transitions and in Ref.~\cite{Bernabeu:2013qea}. 
 In this paper we break the interpretation
of $A_T$ as a test of T purposefully through the choice of different CP tags, and 
the resulting variations in the effective T violation can be used to probe 
the existence of different small effects. 
In particular, we show that with 
specially chosen  ``generalized'' CP tags the dominant 
amplitudes cancel in observables associated with $A_T$, thus yielding
a direct test of weak phase universality, or, alternatively, a 
measurement of differences of penguin pollution in the SM. These differences have
been difficult to quantify~\cite{Bevan:2014iga}, and our procedure gives direct access to them. 
To explicate this, we shall start by revisiting the interpretation of $A_T$.

\section{Interpreting $A_T$}
\label{sec:interpretation}
The combination of
Einstein-Podolsky-Rosen (EPR)
entanglement in the $B{\bar B}$ system from $\Upsilon(4s)$ decay
with the possibility of {\em both} lepton and CP tagging 
(using $J/\psi K_{S,L}$)
allows a near-perfect experimental 
realization of a process and its time-reversal conjugate, making
the measurement of $A_T$ a true
test of time-reversal symmetry. 
The first tag at $t_0$, of CP (or flavor), sets the initial state
of the remaining particle.
Following the formalism of the recent analysis of 
BaBar's measured $A_T$ by Applebaum et al.~\cite{Applebaum:2013wxa}, 
the state assignment of the remaining $B$-meson can be thought of
as an inverse decay at $t_0$ from the
opposite CP (or flavor) tag. 
Figure 1 visualizes this result. 
The inverse decay is realized through EPR entanglement and the decay of
another particle, and Applebaum et al.~state the conditions
under which a nonzero $A_T$ reveals T violation, though,
as we will show, the conditions turn out to be necessary but
not sufficient. That is, they note that (i) the absence of CPT violation in
strangeness changing decays and (ii) the absence of wrong sign decays or
the absence of direct CP violation in semileptonic decays if wrong sign
decays occur are required to interpret
$A_T$ as a test of T invariance~\cite{Applebaum:2013wxa}. (A complementary discussion
of the conditions under which $A_T$ serves as a test of T can be found in 
Ref.~\cite{Bernabeu:2013qea}.) 
Figure 1a illustrates the ideal case in which the detection of one state projects
the other $B$-meson into the state orthogonal to it, thus realizing
the exchange of initial and final states needed to
construct the time-conjugate process.

\begin{figure}
\centering
\includegraphics[width=\textwidth, clip = true, trim = 2mm 16mm 0 0]{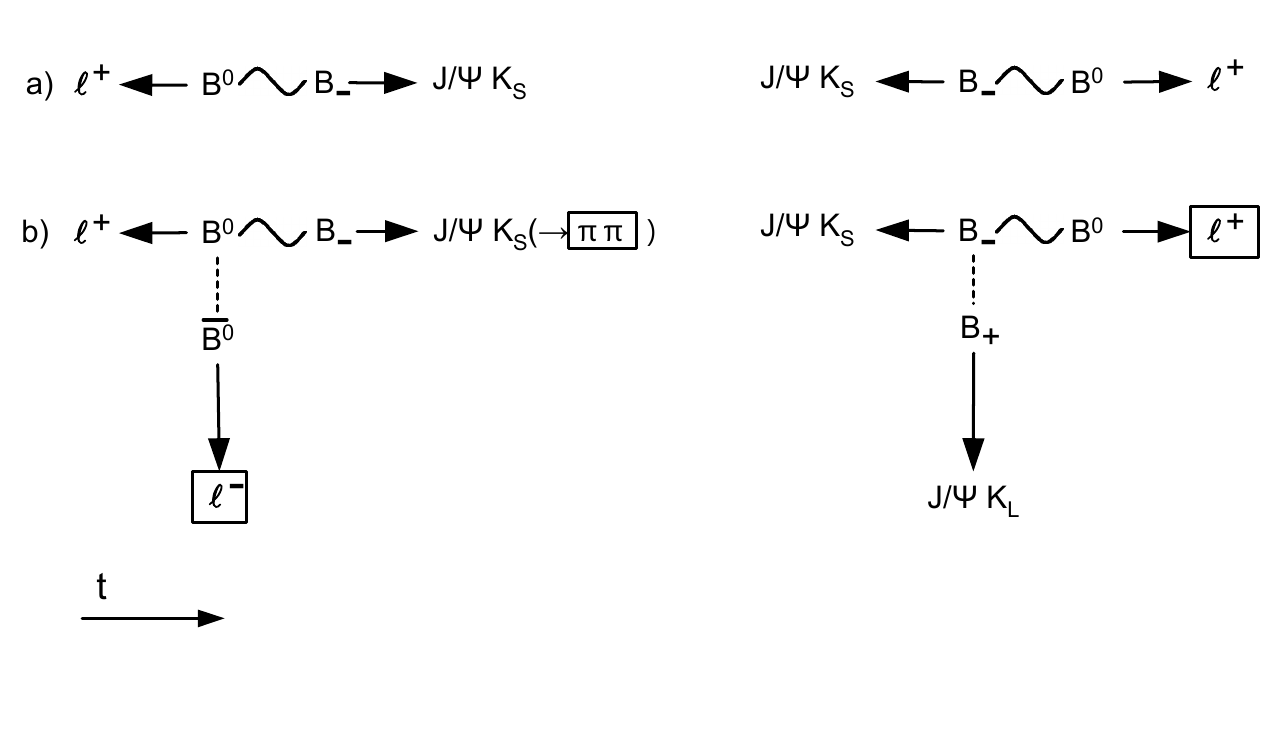}
\caption{The transition
$B^0\rightarrow B_-$
and the construction of its time-conjugate $B_- \rightarrow B^0$.
a)
Idealized:
the initial detection of $\ell^-$ projects the other $B$ into the
orthogonal flavor state, realizing $B^0\rightarrow B_-$ upon subsequent detection
of  $J/\psi K_S$, whereas the initial detection of $J/\psi K_L$
projects the other $B$ into the $\text{CP}=-$ state.
In this latter
case subsequent detection of $\ell^+ X$ realizes $B_- \to B^0$,
the time-reversed process associated with $B^0\to B_-$.
The initial-state projections can be thought of as
inverse decays of $\ell^+$ and $J/\psi K_S$,
respectively~\cite{Applebaum:2013wxa}.
b) Expanded to include the particles that are detected (boxes)
to tag the initial and final states of the $B$-meson. The second process is not
the time conjugate of the first once direct CP violation in the tagging
decay is included. The CP state of the $B$-meson prepared through
inverse decay is not identical to that of the $B$ which decays to
$J/\psi K_{S}(\pi^+\pi^-)$. Note at the $B$-factories that $K_L$ is reconstructed
through its interactions with the detector~\cite{Harrison:1998yr}.
}
\label{fig:TRInterpretation}
\end{figure}

There is one more effect to consider in interpreting $A_T$ as a test of T, 
and it can arise if the CP tagging state is itself 
reconstructed through its decay to hadrons. 
That is, direct CP violation in the decay of CP tag to 
hadronic final states 
breaks the ability to construct the time-reversed process. 
(This is distinct from
the complications due to $\epsilon_K$, noted in Ref.~\cite{Applebaum:2013wxa}.)
Figure 1b illustrates this, though 
the details are provided in the following section. 
Ideally, $K_S$ and $K_L$ can be reconstructed unambiguously, but 
direct CP violation in the reconstruction of the $K_S$ from $K_S\to \pi\pi$ decay 
prevents this. 
In the formalism of \cite{Applebaum:2013wxa}, it appears
as if it were a CPT-violating effect. Of course, 
CPT is not actually broken, but, 
rather, the relationships between the T and CP asymmetries 
expected
under an assumption of CPT invariance will not hold 
because of direct CP 
violation in the kaon decay. 
The effect of direct CP violation in $K_S\to \pi\pi$ is 
numerically very small~\cite{Beringer:1900zz}. Nevertheless it 
can limit the sensitivity of CPT tests that follow 
from comparing T and CP asymmetries, $A_T$ and $A_{CP}$. 
(We note that the best limits on the real part of the CPT-violating parameter $z$ in the 
$B$ system 
comes from studies of $b\to c\bar c s$ decay~\cite{Higuchi:2012kx,Schubert:2014ska}.)
The new method we propose exploits the potential failure of $A_T$ as a test of T
by selecting CP tags of common dominant weak phase (in the SM) but differing
penguin pollution, e.g., 
to yield new observables --- this is illustrated in Fig.~2. These new 
observables 
probe small effects that have not previously been directly measured. 
In these cases
as well we find $|A_T| \ne |A_{CP}|$ without CPT violation.
We now turn to the details.
\begin{figure}
\centering
\includegraphics[width=\textwidth, clip = true, trim = 11.5mm 47mm 8.5mm 0 ]{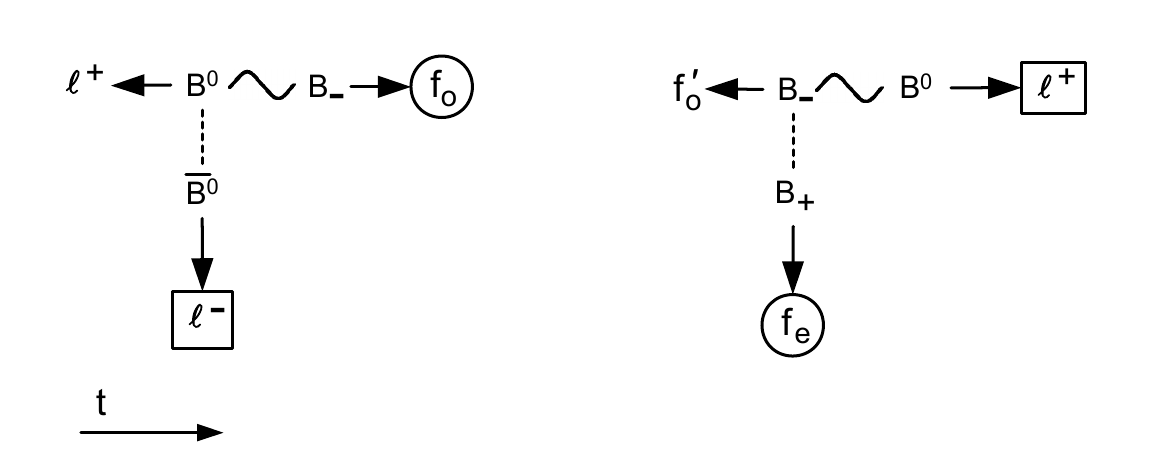}
\caption{ The $B^0\to B_-$ transition and its time conjugate
using general CP tags $f_o$ and $f_e$, which are odd and even, respectively, under
CP. In this case the interpretation of $A_T$
as a test of T can be broken at the tag level. Thus detecting $f_e$ at $t_0$ is tantamount
to the inverse decay $f_o^\prime \rightarrow B_-$, where $f_o$ and
$f_o^\prime$ are distinct states. Here circles
are used to indicate that
the CP tag may be reconstructed rather than directly detected.
}
\label{fig:GeneralCPTags}
\end{figure}

\section{Details}
\label{sec:details}
The time-dependent decay rate for $B \bar B$ mesons
produced in $\Upsilon(4S)$ decay, in which one $B$ decays to
final state $f_1$ at time $t_1$ and the other  decays
to final state $f_2$ at a later time $t_2$ has been analyzed in the
presence of CPT violation, wrong-sign
semileptonic decays, and  wrong strangeness decays~\cite{Applebaum:2013wxa}.
In what follows we assume all of these refinements
to be completely negligible. Moreover,
we neglect CP violation in $B\bar B$ mixing and set the width difference
of the $B$-meson weak eigenstates to zero, i.e., 
$\Gamma_H - \Gamma_L=0$.
The decay rate to $f_1$ and then $f_2$ is denoted as
$\Gamma_{(f_1)_\perp,f_2}$ and is thus given by
\begin{eqnarray}
\label{cpdef}
\Gamma_{(f_1)_\perp,f_2} &=& {\cal N}_1 {\cal N}_2
e^{-\Gamma(t_1+t_2)}[ 1 + C_{(1)_\perp,2}\cos(\Delta m_B\,t)
\nonumber \\
&+& S_{(1)_\perp,2}\sin(\Delta m_B\,t) ]\,,
\end{eqnarray}
with $\Gamma\equiv (\Gamma_H + \Gamma_L)/2$, $\Delta m_B \equiv m_H - m_L$,
$t=t_2-t_1\ge 0$, $S_{(1)_\perp,2} \equiv C_1S_2-C_2S_1$, and
$C_{(1)_\perp,2} \equiv -[C_2C_1+S_2S_1]$~\cite{Applebaum:2013wxa}.
Moreover, $C_f\equiv ({1-|\lambda_f|^2})/({1+|\lambda_f|^2})$ and
$S_f\equiv 2\Im(\lambda_f)/(1+|\lambda_f|^2)$,
where $\lambda_f\equiv (q/p)({\bar A}_f/A_f)$, noting
$A_f \equiv A(B^0\to f)$, ${\bar A}_f \equiv A({\bar B}^0\to f)$,
${\cal N}_f \equiv A_f^2 + {\bar A}_f^2$, and $q$ and $p$ are the usual $B\bar B$
mixing parameters~\cite{Beringer:1900zz}. Since we neglect
wrong-sign semileptonic decay, $C_{\ell^+ X} = - C_{\ell^- X} = 1$.
Defining normalized rates as per
$\Gamma^\prime_{(f_1)_\perp,f_2} \equiv  \Gamma_{(f_1)_\perp,f_2}/({\cal N}_{f_1}{\cal N}_{f_2})$
we have, in the case of the asymmetry illustrated in Fig.~1,
\begin{equation}
A_T = \frac{\Gamma^\prime_{(\ell^- X)_\perp,J/\psi K_S} - \Gamma^\prime_{(J/\psi K_L)_\perp,\ell^+ X}}{\Gamma^\prime_{(\ell^- X)_\perp,J/\psi K_S} +
\Gamma^\prime_{(J/\psi K_L)_\perp,\ell^+ X}}\,.
\label{normasym}
\end{equation}
Note that normalizing each rate is important to
a meaningful experimental asymmetry because the $J/\psi K_S$ (or, more generally, $c{\bar c} K_S$)
and $J/\psi K_L$ final states have
different reconstruction efficiencies~\cite{Bernabeu:2012ab}. BaBar constructs
four different asymmetries, based on four distinct subpopulations of events, namely,
those for $\Gamma_{(\ell^+ X)_\perp, c{\bar c} K_S}$ (${\bar B}^0 \to B_-$),
$\Gamma_{(c{\bar c} K_S)_\perp, \ell^+ X }$ (${B}_+ \to B^0$),
$\Gamma_{(\ell^+ X)_\perp, J/\psi K_L}$ (${\bar B}^0 \to B_+$),
$\Gamma_{(J/\psi K_L)_\perp, \ell^+ X }$ (${B}_- \to B^0$),
and their T conjugates, respectively,
and finds the measurements of the individual asymmetries to be compatible~\cite{Lees:2012uka}. We note that the
normalization factors ${\cal N}_f$ for general CP tags will differ;
nevertheless, meaningful experimental asymmetries can be constructed
through the use of normalized decay rates as already implemented in
BaBar's $A_T$ analysis~\cite{Lees:2012uka}. 

In what follows we generalize the choice of CP final states, so that
$J/\psi K_S \to f_o$ and  $J/\psi K_L \to f_e$, where
``o'' (``e'') denotes a CP-odd (even) final state.
We define
\begin{eqnarray}
\!\!\!\!\!\!A_{CP}^{e+} &\equiv& \frac{\Gamma^\prime_{(\ell^-X)_\perp,f_e}-\Gamma^\prime_{(\ell^+X)_\perp,f_e}}
{\Gamma^\prime_{(\ell^-X)_\perp,f_e}+\Gamma^\prime_{(\ell^+X)_\perp,f_e}} \nonumber \\
&=& C_e \cos(\Delta m_B\,t)-S_e\sin(\Delta m_B\,t)\,,
\\
\!\!\!\!\!\!A_{CP}^{e-} &\equiv& \frac{\Gamma^\prime_{(f_e)_\perp,\ell^-X}-\Gamma^\prime_{(f_e)_\perp,\ell^+X}}
{\Gamma^\prime_{(f_e)_\perp,\ell^-X}+\Gamma^\prime_{(f_e)_\perp,\ell^+X}} \nonumber \\
&=& C_e \cos(\Delta m_B\,t)+S_e\sin(\Delta m_B\,t)\,,
\end{eqnarray}
where $A_{CP}^{e+} \to A_{CP}^{o+}$
and $A_{CP}^{e-} \to A_{CP}^{o-}$
follow by replacing $f_e \to f_o$. Note that
$A_{CP}^{f+}$ and $A_{CP}^{f-}$ employ distinct data samples.
Moreover,
\begin{eqnarray}
\!\!\!\!\!\!A_T^{o+} &\equiv& \frac{\Gamma^\prime_{(f_o)_\perp,\ell^-X}-\Gamma^\prime_{(\ell^+X)_\perp,f_e}}
{\Gamma^\prime_{(f_o)_\perp,\ell^-X}+\Gamma^\prime_{(\ell^+X)_\perp,f_e}} \nonumber \\
&=& \frac{(C_e+C_o)\cos(\Delta m_B\,t)+(S_o-S_e)\sin(\Delta m_B\,t)}
{2+(C_o-C_e)\cos(\Delta m_B\,t)+(S_o+S_e)\sin(\Delta m_B\,t)}\,,\\
\!\!\!\!\!\!A_T^{o-} &\equiv& \frac{\Gamma^\prime_{(\ell^-X)_\perp,f_o}-\Gamma^\prime_{(f_e)_\perp,\ell^+X}}
{\Gamma^\prime_{(\ell^-X)_\perp,f_o}+\Gamma^\prime_{(f_e)_\perp,\ell^+X}} \nonumber \\
&=& \frac{(C_e+C_o)\cos(\Delta m_B\,t)-(S_o-S_e)\sin(\Delta m_B\,t)}
{2+(C_o-C_e)\cos(\Delta m_B\,t)-(S_o+S_e)\sin(\Delta m_B\,t)}\,,
\end{eqnarray}
and 
\begin{eqnarray}
\!\!\!\!\!\!A_T^{e+} &\equiv& \frac{\Gamma^\prime_{(f_e)_\perp,\ell^-X}-\Gamma^\prime_{(\ell^+X)_\perp,f_o}}
{\Gamma^\prime_{(f_e)_\perp,\ell^-X}+\Gamma^\prime_{(\ell^+X)_\perp,f_o}} \nonumber \\
&=& \frac{(C_e+C_o)\cos(\Delta m_B\,t)-(S_o-S_e)\sin(\Delta m_B\,t)}
{2-(C_o-C_e)\cos(\Delta m_B\,t)+(S_o+S_e)\sin(\Delta m_B\,t)}\,,\\
\!\!\!\!\!\!A_T^{e-} &\equiv& \frac{\Gamma^\prime_{(\ell^-X)_\perp,f_e}-\Gamma^\prime_{(f_o)_\perp,\ell^+X}}
{\Gamma^\prime_{(\ell^-X)_\perp,f_e}+\Gamma^\prime_{(f_o)_\perp,\ell^+X}} \nonumber \\
&=& \frac{(C_e+C_o)\cos(\Delta m_B\,t)+(S_o-S_e)\sin(\Delta m_B\,t)}
{2-(C_o-C_e)\cos(\Delta m_B\,t)-(S_o+S_e)\sin(\Delta m_B\,t)}\,.
\label{ATnew}
\end{eqnarray}
Each time-dependent asymmetry has four parameters made
distinguishable by the
various time-dependent functions, and they can be
measured experimentally. Indeed the individual asymmetries can be
simultaneously fit for $S_o+S_e$, $S_o-S_e$, $C_o+C_e$, and
$C_o-C_e$.
Note that if $C_o=C_e$ and $S_o=-S_e$,
$A_{CP}^{e+}= A_{CP}^{o-}= A_T^{o+} = A_T^{e-}$ and
$A_{CP}^{e-}= A_{CP}^{o+}= A_T^{o-} = A_T^{e+}$. 
Neglecting CP violation in kaon decay, we 
note that $\lambda_{J/\psi K_S} = - \lambda_{J/\psi K_L}$. The $K_S$ is reconstructed
through its decays to $\pi\pi$ (2$\pi$), whereas the $K_L$, at BaBar and Belle, is not determined
from its decay to $\pi^0\pi^0\pi^0$, 
though this can be done at DAPHNE~\cite{Bernabeu:2012nu}. We 
calculate $\lambda_{2\pi}$ 
\begin{eqnarray}
\lambda_{2\pi}&=&\frac{q}{p}\frac{\langle\bar K^0|\bar B^0\rangle}{\langle K^0|B^0\rangle}\frac{1+\epsilon_K}{1-\epsilon_K}\frac{1+\eta_{2\pi}}{1-\eta_{2\pi}},
\end{eqnarray}
where $\eta_{2\pi}\equiv{\langle2\pi|K_L\rangle}/{\langle2\pi|K_S\rangle}$ 
and $\epsilon_K$ captures CP violation in $K{\bar K}$ mixing. 
Since $\eta_{2\pi}\neq 0$~\cite{Beringer:1900zz},
we find $C_{2\pi}\neq C_{K_L}$
and $S_{2\pi}\neq -S_{K_L}$, yielding $|A_{CP}|\neq |A_T|$ (in all cases)
without CPT violation. Though we concur with Ref.~\cite{Applebaum:2013wxa} that neither
direct CP violation in $B$ meson decay nor CP violation in $K {\bar K}$
mixing can generate this effect,  we see explicitly that the effect of direct CP violation in $K$ 
decay can be included through a nonzero $\theta_f$, 
a nominally CPT-violating parameter, in the formalism of
Ref.~\cite{Applebaum:2013wxa}. We note the criteria of 
Applebaum et al.~\cite{Applebaum:2013wxa}, enumerated in the previous section,   
should be supplemented with the neglect of direct CP violation 
in kaon decay, if the kaon is reconstructed through its hadronic decays, 
in order to interpret $A_T$ as a test of $T$.

Thus far we have discussed the CP final states $f_o=J/\psi K_S$ and
$f_e=J/\psi K_L$, though other choices are possible. If we choose CP final states
that share a dominant weak phase with each other and with $J/\psi K_{S,L}$,
we have
$f_{o^\prime}=\phi K_S,\eta K_L,\eta^\prime K_L, \rho^0 K_S, \omega K_S, \pi^0 K_L$
and
$f_{e^\prime}=\phi K_L,\eta K_S,\eta^\prime K_S, \rho^0 K_L, \omega K_L, \pi^0 K_S$,
respectively, with the prime notation henceforth representing a CP tag other than $J/\psi K_{S,L}$. These are the two-body ``$\sin(2\beta)$'' modes commonly studied\footnote{Three-body decays, such as $K_S K_S K_S$ or
$K^+ K^- K_S$, have also been studied, though determining the CP
content of the $K^+ K^- K_S$ Dalitz plot requires an angular moment 
analysis~\cite{Gershon:2004tk,Lees:2012kxa}.}
to test its universality~\cite{Beringer:1900zz,Aushev:2010bq}.
Not only can we use these
modes to form the $A_T$
asymmetries we have discussed thus far~\cite{Bevan:2013rpr}, such as the comparison of
${\bar B}^0 \to B_{o^\prime}$ with $B_{o^\prime} \to {\bar B}^0$, we can form
two more for each one: e.g., we can compare
$\bar{B}^0 \to B_{o^\prime}$ to $B_{o} \to \bar{B}^0$, as well as
$\bar{B}^0 \to B_{o}$ to $B_{o^\prime} \to \bar{B}^0$.
Turning to Eq.~(\ref{ATnew}), 
we see that the parameters associated with the
$\sin(\Delta m_B t)$ terms in these comparisons
are, e.g., $S_{o'} - S_e$  and $S_{o'} + S_e$. In $S_{o'} + S_e$ 
the dominant weak
phase contributions (in the SM) cancel, and the small terms, namely, the penguin
contributions, as
well as possible contributions from new physics, are determined directly. 
In the analogous comparison of
$\bar{B}^0 \to B_{e^\prime}$ with $B_{e} \to \bar{B}^0$ decay, the dominant weak phases
cancel in $S_{o} + S_{e'}$. Note that the possibility of a direct 
measurement of a quantity in which 
the dominant weak phases can cancel is special to the $A_T$ construction.

In order to demonstrate this, we first define the 
parameter $\lambda_f$ on which $S_{o,e}$ depend.
There is a factor of $\exp({-i2\beta})$ from $B\bar B$ mixing,
and, in general, the decay amplitude can be written
as a linear combination of 2 weak phases (we select
``up'' and ``charm''): $A_f = a^c_f e^{-i\theta_c} + a^u_f e^{-i\theta_u}$,
in which ``$a^c_f$'', ``$a^u_f$'' contain the magnitudes of the amplitude
associated with each phase, including diagrammatic tree and penguin contributions.
The associated weak phases are $\theta_c = 0$ and $\theta_u \equiv \gamma$.
The dominant weak phase is determined by the quark-flavor content of the
final state. Our focus is on the $\sin(2\beta)$ modes, for which $a^c_f$
is the dominant amplitude. Defining
\begin{eqnarray}
\lambda_f &=& -\eta_{CP}^f e^{-2i\beta}\frac{1+d_fe^{-i\gamma}}{1+d_fe^{i\gamma}}\,,\\
d_f &\equiv& \left|
\frac{V_{ub}^* V_{us}^{}}{V_{cb}^* V_{cs}^{}}\right|\frac{a^u_f}{a^c_f}\,,
\end{eqnarray}
where $\text{CP} |f\rangle = \eta_{CP}^{f} |f\rangle$,
a simple calculation gives us:~\cite{Beneke}
\begin{eqnarray}
S_f &=& -\eta_{CP}^{f}\frac{\sin(2\beta)+2\Re(d_f)\sin(2\beta+\gamma)+|d_f|^2\sin(2\beta+2\gamma)}{1+|d_f|^2+2\Re(d_f)\cos(\gamma)} \,,\nonumber \\
C_f &=& \frac{-2\Im(d_f)\sin(\gamma)}{1+|d_f|^2+2\Re(d_f)\cos(\gamma)}\,.
\end{eqnarray}
As long familiar, a
difficulty arises in attempting to separate the dominant term
from any small effects. Setting the smaller,
wrong phase contribution to zero, we recover
the simplified expressions
$C_f = 0$, $S_f = -\eta_{CP}^f \sin(2\beta)$ for all $f$. It is
convenient to define $\delta S_f$ such that
$S_f = -\eta_{CP}^f (\sin(2\beta) + \delta S_f)\,$.\footnote{We use 
``$\delta S_f$'' in place of the ``$\Delta S_f$'' used in 
Refs.~\cite{Boos:2004xp,Beneke,Cheng,Buchalla,Li:2006vq,Dutta:2008xw,Frings:2015eva}
in order to avoid confusion with the quantities $\Delta S_T^\pm$ 
of Refs.~\cite{Bernabeu:2012ab,Lees:2012uka,Applebaum:2013wxa} that we have already introduced.}

Several 
theoretical  
studies have been made of the
deviations of $S_f$, measured through $A_{CP}^{f+}$,
from $\sin(2\beta)$, through computation of the amplitudes 
in the SM~\cite{Boos:2004xp,Beneke,Cheng,Buchalla,Li:2006vq,Dutta:2008xw,Frings:2015eva}, 
as well as through approaches using 
SU(3)-flavor-based assumptions~\cite{Grossman:2003qp,Jung:2012mp,Ligeti:2015yma}. 
A particular effort has been placed on determining the size of the small penguin pollution in 
the golden $J/\psi K_{S,L}$ modes, for which ancillary data and flavor-based
relations can be 
used~\cite{Fleischer:1999nz,Ciuchini:2005mg,Faller:2008zc,Ciuchini:2011kd,Ligeti:2015yma}. 
Experimentally one can form
\begin{eqnarray}
\delta S_f = -\eta_{CP}^f S_f - \sin(2\beta)
\label{usual}
\end{eqnarray}
using the determination of $\sin(2\beta)$ in
$B\to c{\bar c}\, K_S$ 
and $J/\Psi K_L$ final states ~\cite{Adachi:2012et,Lees:2012uka,Aaij:2015vza}, 
though the error in $\delta S_f$ is dominantly
that in $S_f$.
We now compare this
procedure to our $A_T$ method
with generalized CP tags.
In this new case, assuming $\sin(2\beta)$ universality,
the $\sin(2\beta)$ term in $S_f$ cancels, yielding
\begin{eqnarray}
(S_e + S_o) = \delta S_o - \delta S_e
\end{eqnarray}
and providing a direct measurement of the difference
of deviations from $\sin(2\beta)$
for the chosen CP tags. If we use a golden mode 
for which $\delta S_{e(o)}\approx 0$, such as $J/\Psi K_{S,L}$, 
to define $\sin(2\beta)$,
then $S_e+S_o \approx \pm \sin (2\beta_{o(e)}) \mp  \sin (2\beta) \pm
\delta S_{o(e)}$, where the upper sign is associated with $o$. 
Thus we test the deviation of $S_f$ from $\sin(2\beta)$ through a single asymmetry measurement, 
whereas a ``double'' difference appears in Eq.~(\ref{usual}). Of course
$\sin(2\beta)$ in $B\to c{\bar c}\, K_S , \: J/\Psi K_L$ decays is 
very well known ($0.677 \pm 0.020$ \cite{Bevan:2014iga}),
so that it is more pertinent to note that the asymmetry $A_T$
can directly employ these highly precise
decay samples as well~\cite{Adachi:2012et,Lees:2012uka,Aaij:2015vza}. 

An asymmetry $A_T$ generally requires the comparison
of the rates ($(\ell^{\pm} X)_{\perp} ,f_{o(e)}$) and ($ (f_{e'(o')})_{\perp},
\ell^{\pm} X$), or of their time conjugates,
while $A_{CP}$ only requires the comparison of
the ($(\ell^{\pm}X)_{\perp},f_{o'(e')}$) rates. Thus in the case of
$\eta^\prime K_S$, e.g., the determination of $S_{e'}$ via $A_{CP}$ employs two
subsamples of limited statistics, whereas the determination of
$S_{e'} + S_o$ via $A_{T}$ is formed from the comparison of a limited
statistics sample 
with the plentiful statistics of $c{\bar c} K_S$. Consequently, we expect
improved access to $\delta S_{e'}$, for any of the CP-even modes that probe
$\sin(2\beta)$, and analogous improvements to the determination
of $\delta S_{o'}$ for any of the CP-odd modes. Current experimental
results for $S_f$ have limited precision in many of the $\sin(2\beta)$ modes
previously listed as CP-tag candidates (e.g.,
$-\eta_{CP}^f S_{\pi^0 K_S} = 0.57 \pm 0.17$;
$-\eta_{CP}^f S_{\omega K_S} = 0.45 \pm 0.24$ \cite{Bevan:2014iga}).
Our method will be of greatest impact for these more poorly known modes. 
Comparing these results against predicted values of 
$\delta S_{o'(e')}$ in the SM should then yield sharper tests of
new physics. 
Such sharpened determinations should also improve the ability to extract the true value of 
$\sin(2\beta)$ 
from fits to the experimental results in a theoretical framework including leading SU(3)
flavor-breaking effects~\cite{Jung:2012mp}, again leading to improved tests of new physics. 
We note that diverse sources of the latter have been 
proposed~\cite{Grossman:1996ke,Hiller:2002ci,Burdman:2003nt,Buras:2004ub,Buchalla}. 

Our method requires the construction of normalized subsample rates as
in Eq.~(\ref{normasym}); normalized subsample rates have 
already been employed in BaBar's
$A_T$ analysis~\cite{Lees:2012uka}. The efficacy of this procedure
can be roughly assessed through the comparison of BaBar's claimed
significance for the observation of T and CP violation through the
measurement of $A_T$ and $A_{CP}$, respectively. In this
exact case BaBar measures  T violation at 14$\sigma$ and CP violation 
at 17$\sigma$~\cite{Lees:2012uka}, so that they are not very different, 
particularly when one notes that the $A_T$ measurement
employs a $J/\psi K_L$ subsample as well. Consequently, for
various $f_{o'(e')}$ we can expect a sharper determination of
$\delta S_{o'(e')}$ through the measurement of $A_T$ than possible
through study of $A_{CP}$ alone.

The method we have proposed can be generalized to other sorts of decay modes, 
such as those that probe $\sin(2\alpha)$~\cite{GY2}. 
The basic idea is that the CP-tagging modes are chosen so that their 
dominant decay amplitudes (in the SM) share the same weak phase. 
In the cases we have considered in this paper, the CP-even and odd tags 
are chosen with a common dominant weak phase of $\sin(2\beta)$. 
In so doing, $A_T$ is no longer a true test of T, but we 
introduce new observables that permit a direct measurement of
small departures from weak-phase universality. If the dominant weak phase 
is universal, then these observables measure 
the penguin pollution in these decays. We emphasize that although the phrase ``penguin trapping''
has previously been used to refer to the specific reconstruction of the penguin
amplitude using flavor-based assumptions and empirical data~\cite{Lipkin:1991st}, 
we use it here to refer to a method by which 
a more precise empirical assessment can be made of observables in which penquin effects can 
appear. 

\section{Summary}
We have described how a broader measurement program of the time-dependent
asymmetry $A_T$ with generalized CP tags, possible at a B factory, 
can be used to measure small
departures from weak-phase universality.
Generally
an analysis of $A_T$ provides four parameters
composed of linear combinations of $S_{o(e)}$ and $C_{o(e)}$; under the
use of generalized CP tags the asymmetry $A_T$ no longer serves as
a genuine T test --- and $|A_T| \ne |A_{CP}|$ can appear without CPT violation.
However, the new observables the $A_T$ construction offers
allow the direct measurement of 
the penguin effects with improved statistical control,
information that can be used to test the universality of $\sin(2\beta)$.
New results of greater precision 
can be obtained from existing B-factory data using this method,
and we believe it can also greatly enable precision studies of CP violation
anticipated with the Belle II detector at KEK.

\section*{Acknowledgments}
We thank Helen Quinn for helpful suggestions regarding the presentation of our paper, 
and we thank Klaus Schubert for a discussion of CPT tests in the B-meson system. 
We acknowledge partial support from the U.S. Department of Energy Office
of Nuclear Physics under contract DE-FG02-96ER40989.


\end{document}